\documentclass[12pt,letterpaper]{JHEP}

\usepackage{epsfig}

\def\gsim{\;\raise0.3ex\hbox{$>$\kern-0.75em\raise-1.1ex\hbox{$\sim$}}\;}
\def\lsim{\;\raise0.3ex\hbox{$<$\kern-0.75em\raise-1.1ex\hbox{$\sim$}}\;}

\def\pref#1{(\ref{#1})}
\def\ignore#1{}  
\def\roughly#1{\mathrel{\raise.3ex\hbox{$#1$%
\kern-.75em\lower1ex\hbox{$\sim$}}}}
\def\lsim{\roughly<}
\def\gsim{\roughly>}
\def\pbrane{$p$-brane}
\def\eq{\begin{equation}}
\def\eeq{\end{equation}}
\def\eqa{\begin{eqnarray}}
\def\eeqa{\end{eqnarray}}
\def\ms{M_s}
\def\mpl{M_p}
\def\rmd{{\rm d}}
\def\dperp{{d_\perp}}
\def\rp{{r_\perp}}
\def\rpl{{r_\parallel}}

\def\vpl{{V_\parallel}}

\title{The Inflationary Brane-Antibrane Universe}

\author{C.P. Burgess,$^{1,2}$ M. Majumdar,$^3$ D. Nolte,$^1$ 
F. Quevedo,$^3$ G. Rajesh,$^1$ R.-J. Zhang$^1$\\

          $^1$ School of Natural Sciences,
               Institute for Advanced Study,\\
               Princeton NJ 08540 USA.\\
          $^2$ Physics Department, McGill University,
               3600 University Street,\\
               Montr\'eal, Qu\'ebec,  H3A 2T8 CANADA.\\
          $^3$ Centre for Mathematical Sciences, DAMTP, 
               University of Cambridge,\\
               Cambridge CB3 0WA UK.\\
          }

\abstract{
We show how the motion through the extra dimensions of a gas of branes
and antibranes can, under certain circumstances, produce an era of
inflation as seen by observers trapped on a 3-brane, with the inflaton
being the inter-brane separation.  Although most of our discussion
refers to arbitrary \pbrane s, when we need to be specific we assume
that they are $D$-branes of Type~II or Type~I string theory.  For
realistic brane couplings, such as those arising in string theory, the
inter-brane potentials are too steep to inflate the universe for
acceptably long times. However, for special regions of the parameter
space of brane-antibrane positions the brane motion is slow enough for
there to be sufficient inflation. Inflation would be more generic in
models where the inter-brane interactions are much weaker.  The
spectrum of primordial density fluctuations predicted has index $n$
slightly less than~1, and an acceptable amplitude, provided that the
extra dimensions have linear size $1/r \sim 10^{12}$ GeV. Reheating
occurs as in hybrid inflation, with the tachyonic instability of the
brane-antibrane system taking over for small separations. The tachyon
field can induce a cascade mechanism within which higher-dimension
branes annihilate into lower-dimension ones. We argue that such a
cascade naturally stops with the production of 3-branes in
10-dimensional string theory.
}

\preprint{McGill-01/09, DAMTP-2001-40, hep-th/0105204}
\dedicated{Dedicated to the memory of Detlef Nolte}

\begin{document}

\section{Introduction and Summary}

Inflationary cosmology~\cite{Guth} provides a compelling
explanation for many otherwise unnatural features of the observed
early universe.  However, no equally compelling microscopic model of the
dynamics underlying inflation has yet emerged. All extant models seem
to require fine-tuning in order to reproduce all of the features which
successful inflation must have: sufficient universal expansion, an
acceptable density spectrum and amplitude of primordial density
fluctuations, and so on~\cite{Rev}.

The relatively recent realization that spacetime may be full of
\pbrane s---or ($p+1$)-dimensional defects---on which most or all of
the observed particle species may be trapped~\cite{braneworld},
provides a new framework within which to search for natural mechanisms
for inflation. Indeed, for many choices of parameters (like the
dimension of space, and the energy scale of brane physics)
inflationary cosmology may provide one of the few constraints we have on
the various possible brane constructions.  This has inspired several
investigations of how inflationary ideas might be rooted in the
brane-world framework~\cite{branecosmology,BPSpheno,Ekpyrotic}.

In this paper we examine the possibility that inflation can be due
to the motion of branes and antibranes\footnote{An antibrane
is an extended object which has the same tension as the corresponding brane,
but with opposite Ramond-Ramond (R-R) charge.} within the extra dimensions, with inflation
ending when branes and antibranes collide and annihilate. Other scenarios
were considered before in terms of only branes, without antibranes. These
earlier analyses either made phenomenological assumptions about the
nature of brane-brane interactions~\cite{BPSpheno}, or were performed
assuming the collisions of BPS branes in M theory~\cite{Ekpyrotic},
for which the interactions are known to vanish to all orders in
perturbation theory and so for which some form of non-perturbative
potential must be assumed.

Brane-antibrane collisions are natural candidates for inflation, since
branes and antibranes interact with one another in a way which has
long been well understood at large distances~\cite{Polchinski}, and
since their annihilation might be expected to provide a relatively
efficient mechanism for converting the energy of relative brane motion
into reheating the universe. Furthermore, much progress has been made
of late in understanding their interactions at short distances within
string theory~\cite{Tachyon}.  (See~\cite{Alexander} for a discussion
of some inflationary issues associated with short-distance brane
interactions.)

In the calculations to be presented here, we explore inflation due to
brane-antibrane collisions for branes of arbitrary dimension, using
what is known about their long- and short-distance interactions. 
In a nutshell, our main findings are these:
\begin{itemize}
\item We find that if the brane-antibrane couplings are very weak then
their motion towards one another is slow enough for sufficient
inflation to occur for generic initial brane positions. Unfortunately,
the couplings derived from string theory are too strong for this
mechanism to work for generic initial conditions, since the branes and
antibranes collide too quickly to provide sufficient
inflation. Inflation of this type, if it occurs, gives rise to a
spectrum of primordial density fluctuations with a spectral index $n
\lsim 1$.
\item For string theory couplings inflation can nonetheless arise, but
only if the initial brane-antibrane configuration starts near special
parts of the extra dimensions where their interactions are
particularly weak. We identify these special initial conditions
explicitly for toroidal compactifications and show that they are
sufficiently large for inflation not to be unlikely given a reasonable
density of branes and antibranes in the extra dimensions.  Primordial
density fluctuations are also predicted to have $n \lsim 1$, and have
the correct amplitude in this scenario if the compactification scale
is of order $1/r \sim 10^{12}$ GeV.
\item Once the brane and antibrane get sufficiently close to one
another they very naturally enter a hybrid inflation
regime~\cite{hybridinflation}, where the slow change in inter-brane
separation becomes replaced with a quick roll down a tachyonic
direction. (We have checked that the tachyonic roll is too quick to
provide sufficient inflation on its own, without the inter-brane
relative motion.) This provides an efficient release of the energy of
relative motion into reheating.
\item Brane-antibrane collisions do not always annihilate to a
no-brane final state, so the energy released by reheating can easily
have significant portions trapped on the daughter brane. This is a
good feature since we imagine all standard-model particles to be so
trapped, and so the reheat energy can plausibly find its way into our
observable universe.  Typical reheat temperatures are of order
$10^{13}$ GeV.
\item Since brane-antibrane annihilation typically produces branes of
lower dimension, it would be natural to have a cascade of
annihilations where higher dimensional branes successively annihilate
into lower ones. Furthermore, the endpoint of such a cascade may 
furnish a dynamical explanation for the present-day conditions which
are widely assumed within brane scenarios. For instance, since 
larger-dimension branes can find one another more easily and annihilate, 
the cascade scenario qualitatively suggests a migration of the brane
abundance towards lower-dimensional branes as the universe evolves. 
Furthermore, branes which are $(3+1)$-dimensional (or less) can plausibly
fail to find one another throughout the history of the universe in a 
10-dimensional theory, potentially providing an explanation within string 
theory for why the late universe might prefer to be populated by 3-branes.
Finally, these final survivors of brane collisions can be argued to 
prefer to be aligned parallel with the large dimensions of space, as is
usually assumed in phenomenological applications. 
\end{itemize}

We now turn to the more detailed explanation of these results.

\section{Inflation Due to Inter-Brane Motion}

In this section we present formulae for the predictions of slow-roll
inflation, where the slowly rolling inflaton is the separation between
two branes, $y^\mu = x_1^\mu - x_2^\mu$. After first outlining our
assumptions concerning the brane and bulk dynamics, and the nature of
the interaction between branes, we describe two scenarios in which
inflation might arise. Although both can provide viable inflation if
the strength of the inter-brane interaction is allowed to be
sufficiently weak, only the second is viable (see Sec 2.5) if the
coupling strengths are taken from string theory.

\subsection{The Effective Action} 

We take as our starting point the following low-energy effective
action governing the inter-brane interactions: $S = S_B + S_{b1} +
S_{b2}$, where the bulk action is\footnote{Our metric has signature
$(-+\dots+)$ and we use Weinberg's curvature conventions.}
\eq
\label{BulkAction}
S_B = - \int\rmd^4x\,\rmd^dy\,\sqrt{-g}\,
\left[\,{\ms^{2+d} \over 2} \, e^{-2\phi} \; R\,+\,\cdots \,
\right]\,.
\eeq
Here $\phi$ is the dilaton\footnote{For non-string applications
$\phi=0$ can be taken everywhere, although we shall find that $\phi$ drops
out of most formulae.}, $R$ is the full ($4+d$)-dimensional scalar curvature,
and Eq.~\pref{BulkAction} defines the fundamental (string) scale, $\ms$. 
(For string theory applications, $d=6$.) The ellipses
denote all other bulk massless fields which 
mediate forces between branes.

The dynamical variables of the brane actions are the brane positions,
$x^\mu_i$, with $i = 1,2$ and $\mu = 0,\dots,3+d$. The relevant part
of the action is:
\eq
\label{BraneAction}
S_{bi} = - \int\rmd^4x\,\rmd^{p-3}y\, \sqrt{-\gamma} \; \Bigl[
\,T_p\, +\,  \cdots \, \Bigr]\,,
\eeq
where $\gamma_{ab} = g_{\mu\nu}\,\partial_a x_i^\mu\,\partial_b
x^\nu_i$ is the induced metric on the brane, and $ T_p = \alpha \,
\ms^{p+1} \, e^{-\phi}$ is the brane tension, where $\alpha$ is a
dimensionless constant which drops out of most of our results.

For our purposes we will assume the branes to be parallel, and choose
coordinates for which the branes lie along coordinate surfaces.  If we
then separate the part of $x_i^m$ transverse to the brane into the
relative motion, $y^m = (x_1 - x_2)^m$, of the branes and the motion
of their center-of-mass, $\overline{x}^m = (x_1+x_2)^m/2$, then
expanding in powers of $\partial_a y^m$ gives
\eq
\label{BA1}
S_{b1}\,+\,S_{b2} = - \int\rmd^4x\,\rmd^{p-3}y 
\,\sqrt{-\gamma}\;T_p\,\left[\,2\,+\,  
\frac14\, g_{mn}\, \gamma^{ab}\,\partial_a y^m\, \partial_b y^n \,
+\, \cdots \,\right]\,.
\eeq

\subsection{Inter-Brane Interactions} 

We assume the dominant interaction between the branes at large
distances (compared to $1/\ms$) to be due to the exchange of massless
bulk modes, including the metric, dilaton and any other massless
bosons which couple to the branes. Any such interaction gives an
inter-brane potential energy (per unit brane area) which falls with
separation like $1/y^{\dperp-2}$ where \eq \dperp = d-p+3\,, \eeq is
the number of spatial dimensions transverse to the branes. In most
models the couplings of these modes is gravitational in strength, so
we parameterize the long distance potential energy by
\eq
\label{PEBrane}
{E \over A_p} = -\,\beta\left(  {e^{2\phi} \over \ms^{2+d}} 
\right) \;
{T_p^2 \over y^{\dperp-2}}\,,
\eeq
where $\beta$ is a model-dependent dimensionless number which
characterizes the overall strength of the inter-brane force, and
depends on the details of the number and kinds of bulk states which
are exchanged.

By far the best motivated candidate for these branes are the
$p$-dimensional BPS Dirichlet branes ($Dp$-branes) of Type II string
theory~\cite{Polchinski}.  In string theory, the potential for the
brane-antibrane ($Dp$--$\overline{Dp}$) system is attractive, since
the contributions due to dilaton/graviton exchange are equal to and
reinforce the attraction due to the opposite R-R charges carried by these
branes. This is to be contrasted with the $Dp$--$Dp$ case where the
dilaton/graviton contributions exactly cancel the repulsion due to the
like-sign R-R charges.  Direct calculation~\cite{Polchinski} gives
\eq
\label{kdef}
\beta =   \pi^{-\dperp/2} \Gamma\left[ \frac{\dperp-2}{2} \right]\,.
\eeq

\subsection{The Effective 4D Inflaton} 

Several assumptions are required in order to analyze the implications 
of these interactions for inflating the 4 non-compact dimensions. Our
most crucial assumption is that the moduli of the compact dimensions
are stabilized by some unknown physics. This
assumption precludes the dynamics of these moduli---including in
particular the space's overall volume---from playing a role in
inflation, allowing us to use the volumes 
$V_\perp \equiv r^{d\,-\,p\,+\,3}_\perp$ 
(transverse to the brane) and $V_\parallel \equiv r^{p\,-\,3}_\parallel$ 
(along the brane) as parameters which
we may adjust to our own ends.

The volumes $V_\perp$ and $V_\parallel$ may be freely adjusted,
subject to two constraints. First, their product is related to
the four-dimensional reduced Planck mass, $\mpl = (8\pi
G_N)^{-1/2} \sim 10^{18}$ GeV, by
\eq
\label{Mpdef}
\mpl^2 = e^{-2\phi} \, \ms^{2+d} \, V_\perp \, V_\parallel
\eeq
as is seen by dimensionally reducing the bulk action, 
Eq.~\pref{BraneAction}. We use this expression in what follows
to eliminate the volume $V_\parallel$ in favor of $\mpl$, leaving 
only $V_\perp$ as an adjustable parameter. 

The second constraint we demand of these volumes is
that they be large when measured in units of $\ms$: $\ms\,\rp \gg 1$,
$\ms\,\rpl \gg 1$. For instance, 
this is required within string theory 
in order to be able to treat the bulk dynamics using only the
low-energy effective field theory corresponding to the massless
string states \footnote{This
is also the condition under which the nonlinear contributions of 
Einstein's equations are negligible when considering the gravitational
field of a single $D$-brane, say. Therefore our approximate 
4D effective field theory treatment is 
self-consistent. }

In what follows we find that the amplitude of primordial 
density fluctuations depends only on the combination 
$\mpl\,\rp$ (and not separately on $e^\phi$ or $\ms$, for example), 
with agreement requiring $\mpl\,\rp \sim 10^6$. Since weak-coupling
in string theory requires $\ms < \mpl$, we see that $\ms$ may
easily be chosen to satisfy the constraints $1/\rp \ll \ms < \mpl$.

With these assumptions, we dimensionally reduce to recast the dynamics
of the inter-brane separation, $y$, as that of a canonically
normalized four-dimensional scalar field, which we denote by
$\psi$. (It is effectively a scalar field because only one component of
$y$ is dynamical if the brane and antibrane move toward each other
along a straight line.) Inspection of the brane action, Eq.~\pref{BA1},
shows the canonically normalized field to be related to $y$ by:
\eq
\label{yScaling}
\psi  = y\,\sqrt{{\,T_p\vpl\over2}}
=y\,\left[{\alpha e^{\phi} \over 2
(\ms\rp)^{\dperp}} \right]^{1/2} \ms\,\mpl\,.
\eeq

The effective scalar potential for $y$, obtained by combining
the tension term of Eq.~\pref{BA1} with the inter-brane potential
energy of Eq.~\pref{PEBrane}, becomes: 
\eq
\label{Potdef}
V(y) = A -  {B \over y^{\dperp-2}},
\eeq
with 
\eqa
\label{ABdefs}
A &=& 2T_p\vpl = {2\alpha e^{\phi} \over (\ms\rp)^\dperp} 
\; \ms^2\mpl^2, \nonumber\\
B &=& {\beta e^{2\phi}\over\ms^{2+d}}\,T_p^2\,\vpl = 
{\alpha^2\beta e^\phi\mpl^2\over 
\ms^{2(\dperp-2)}\,r^\dperp_\perp}\,.
\eeqa

These expressions for $V(y)$ apply if the separation between the branes
is large compared to $1/\ms$, but is not large enough to be
comparable to the overall size of the compact transverse
dimensions. In our application to string theory, we shall find that 
in order
to obtain sufficient inflation, one requires inter-brane separations
which are comparable with $\rp$. We therefore require the
modifications to $V$ which arise in this case.

Although we cannot provide explicit expressions for the modified 
$V$ in general
geometries, we can do so for simple cases like tori. The potential
energy of a brane on a torus can be represented by summing the above
potential over a lattice of ``image'' branes. As a brane is moved
to an antipodal point from the source brane on a torus, it can
interact equally strongly with two or more of the image charges.
Inflationary calculations in this instance may be performed using the
potential of Eq.~\pref{Potdef}, but with the interaction term summed
over the contributions of all images.

The potential~\pref{Potdef} neglects several effects. First, it
neglects dependence on the inter-brane speeds and orientations. The
velocity dependence is legitimate for a slow roll, where speeds are
assumed to be initially very small. Neglect of the dependence on
orientation presumes the branes and antibranes to be moving strictly
parallel to one another, and has been done for simplicity of
analysis.
 We believe further study of the dependence of inflation on
these variables to be warranted \footnote{See next section for a 
short discussion of these points.}.

The inter-brane potential also neglects angular momentum terms, 
corresponding to the assumption that the branes of interest are
moving directly towards one another. Since the angular momentum, $L$,
contributes to the radial potential as $V_L \sim L^2/y^2$, we
see that it dominates the large-$y$ limit unless there are three
transverse dimensions (corresponding to $d-p = 0$, {\it e.g.}
$6$-branes in 10 spacetime dimensions). We believe
the neglect of angular momentum to be justified for the branes which
initiate inflation, since these branes must move very slowly. Branes
which attract and which start initially essentially from rest should
approach each other directly.

\subsection{The Case $\ms^{-1} \ll y \ll \rp$} 

Let us first consider the case of a parallel $p$-brane and antibrane
separated from each other by a distance $y$. We assume that the branes
wrap a $(p-3)$-dimensional space of volume $V_{\parallel}$, and that
the space transverse to the branes is compactified on a space of
volume $V_{\perp}$.  We will show that it is only possible to satisfy
the slow-roll conditions for inflation in this set-up if the constant
$\beta$ parameterizing the strength of the inter-brane force is chosen
sufficiently small. The $\beta$ predicted for $Dp$--$\overline{Dp}$ branes
in string theory (see Eq.~\pref{kdef}) turns out to be too large to
permit slow-roll inflation, precluding the generation of inflation
using this mechanism in string theory.

We use the potential of Eq.~\pref{Potdef} which is valid when the
inter-brane separation is large compared to $1/M_s$, but small
compared with the scale $\rp$ of the compact transverse dimensions.
The standard equations of motion of a Friedmann-Robertson-Walker
universe are
\begin{eqnarray}
&&\ddot\psi\,+\,3H\dot\psi = -\,{\rmd V\over\rmd\psi}\,,\nonumber\\
&&H^2 = {1\over 3\mpl^2}\left[V+{\dot\psi^2\over2}\right]\,,
\end{eqnarray} 
where $H$ is the Hubble parameter.  The slow roll conditions require
the friction and potential terms dominate, {\it i.e.}, $|\ddot\psi|\ll
3H|\dot\psi|$ and $\dot\psi^2\ll V$. Consistency with the field
equations requires small values for the two slow-roll parameters
\eq
\epsilon = {\mpl^2\over2}\left({V'\over V}\right)^2\,,
\qquad\qquad
\eta = \mpl^2\,{V''\over V}\,,
\eeq
where the primes denote derivatives with respect to $\psi$.  For
slow-roll inflation, both $|\epsilon|$ and $|\eta|$ should be much
less than $1$.

In our case, $|\epsilon|$ is always smaller than $|\eta|$ and will be
neglected in the following discussion. The $\eta$ parameter is
\eq
\label{etaresult}
\eta \approx -\beta \, (\dperp-1)(\dperp-2)  
\left({\rp\over y}\right)^\dperp\,.
\eeq
Here we see the requirement for small $\beta$. For the string
theoretic value of $\beta$ as in Eq.~\pref{kdef}, the slow-roll
condition $|\eta|\ll 1$ would require a brane separation $y$ much
bigger than the size of transverse dimension $\rp$, which violates the
assumptions under which Eq.~\pref{etaresult} was obtained.\footnote{On
the last point, we disagree with Ref.~\cite{DvaliShafi}.
\label{footnote}}

One might imagine attempting to satisfy the slow-roll condition using
$\beta$ from string theory in Eq.~\pref{etaresult} by using a
compactification manifold which is asymmetric, with some (say, $n$)
transverse dimensions much smaller than the others. This permits
$y > \rp$ if the branes
are separated along one of the long directions.  This stratagem turns out
to fail, however, although the analysis which shows this requires the
inclusion of the corrections to the inter-brane potential which arise
when $y \sim \rp$, such as we consider in the next section. The
potential which results may be shown by a simple scaling argument---or
by considerations of T-duality---to be modified in this case to $V
\sim {1/a^n y^{\dperp-n}}$ (where $a$ is the ``small''
radius). Repeating the calculation for this potential gives $y \gg
\bigl({V_{\perp}\over a^n}\bigr)^{1/(\dperp-n)}$, which is no
improvement over Eq.~\pref{etaresult}.
 
If we adopt at this point a phenomenological approach, and simply
treat $\beta$ as a model-dependent parameter, we may compute the
predictions of the resulting slow-roll inflation.\footnote{Potentials 
similar to Eq.~\pref{Potdef} were considered in~\cite{riotto}
with $B<0$, inspired by nonperturbative supersymmetric potentials. 
The fact
that $B<0$ was the reason they 
found a blue spectrum, in contrast with our
result.} The number of
$e$-foldings occurring after the scales probed by the COBE data leave
the horizon can be computed as
\eq
N = \int^{\psi_{\rm end}}_{\psi_*} {H\over\dot\psi}\,\rmd\psi 
\approx {(y_*/\rp)^\dperp\over \dperp\,(\dperp-2)\,\beta}\,,
\eeq
where $y_*$ is the brane separation when the primordial
density perturbation exits the de Sitter horizon during inflation. We
assume $y_* \gg y_{\rm end}$ in deriving this result. The number of
$e$-foldings can be made sufficiently large ($N\gsim 60$) by choosing
small enough $\beta$.

The amplitude of this density perturbation when it re-enters the
horizon, as observed by Cosmic Microwave Background (CMB) experiments,
may be written as follows:
\eq
\delta_H\approx {1\over 5\,\sqrt{3}\,\pi}\,{V^{3/2}\over \mpl^3\,V'}
\approx {2\over 5\,\sqrt{3}\,\pi}\,
{\left(\dperp\,N\right)^{(\dperp-1)/\dperp}
\over \left[\,(\dperp-2)\,\beta\,\right]^{1/\dperp}}\,
{1\over \mpl\,\rp}\,,
\eeq
where the COBE results imply $\delta_H = 1.9 \times 10^{-5}$~\cite{COBE}.

The spectrum of perturbations is similarly calculated.  For a
(Gaussian) perturbation $\delta({\vec x},t)$, it is conventional to
Fourier transform it into the momentum space, $\delta({\vec
x},t)=\int\rmd^3{\vec k}\,\delta({\vec k},t) \,\exp({\rm i}{\vec
k}\cdot{\vec x})$, where $\vec{k}$ is the wave vector in the comoving
frame and $k$ is its norm.  The power spectrum $P_\delta(k)$ is then
defined in terms of the two-point correlation functions
$\langle\,\delta({\vec k})\,\delta({\vec k}')\,\rangle =\delta^3({\vec
k}-{\vec k}')\,(2\pi^2/k^3)\,P_\delta(k)$.  The function $P_\delta(k)$
has the form $\sim\,k^{n-1}$ where $n$ is known as the spectral
index. In these conventions, inflationary models predict a scale
invariant Harrison-Zel'dovich spectrum ($n=1$) to high precision. In
our case the spectral index and its derivative can be readily
calculated
\eq
n-1\approx -\,{2\,(\dperp-1)\over\dperp\,N}\,,\qquad\qquad
{\rmd n\over\rmd\ln k}\approx-\,{2\,(\dperp-1)\over\dperp\,N^2}\,.
\eeq

\subsection{The Case $y \sim \rp$ and Special Points}

In this subsection we study the case where the brane-antibrane
separation is comparable to the size of compactified transverse
dimensions.  We will show that the difficulty encountered in the
previous case is resolved by an interesting cancellation in the brane
potential.

For our purposes, we assume the compactified transverse manifold to be
a $\dperp$-dimensional square torus with a uniform circumference
$\rp$.  When the brane-antibrane separation is comparable to $\rp$, we
have to include contributions to the potential from $p$-brane images,
{\it i.e.}, we have to study the potential in the covering space of
the torus, which is a $\dperp$-dimensional lattice, $({\bf
{R/Z}})^\dperp$.

The potential at the position of the antibrane is 
\begin{equation}
V({\vec r})\,=\,A\,-\,\sum_{i}\,{B\over|{\vec r}-{\vec r}_i|^{\dperp-2}}\,,
\end{equation}
where the parameters $A$ and $B$ are defined in Eq. \pref{ABdefs},
$\vec r$ and ${\vec r}_i$ are the vectors denoting the positions of
the $p$-branes and antibranes in the $\dperp$-dimensional coordinate
space, and the summation is over all the lattice sites occupied by the
brane images, labelled by $i$.  We schematically show our set-up in
Fig.~\ref{figure1}.

It is easy to show the first and second derivatives of the above
potential vanish when the antibrane is located at the center of a
hyper-cubic cell, which we denote as ${\vec r}_0$. To see the
vanishing of these quantities, compute
\begin{eqnarray}
\left.{\partial\,V\over \partial\,r_a}\right|_{\,{\vec r}={\vec r}_0}
&=&(\dperp-2)\,
\sum_i\,{B\,(r_0-r_i)_a\over|{\vec r}_0-{\vec r}_i|^\dperp}\,,
\nonumber\\
\left.{\partial^2V\over \partial\,r_a\,\partial\,r_b}
\right|_{\,{\vec r}={\vec r}_0}
&=&(\dperp-2)\,\sum_i\,{B\over |{\vec r}_0-{\vec r}_i|^\dperp}
\,\left[\,\delta_{ab}-\dperp\,{(r_0-r_i)_a\,
(r_0-r_i)_b\over|{\vec r}_0-{\vec r}_i|^2}\,\right]\,,\qquad
\label{VDD}
\end{eqnarray}
where $a, b\in \{1,2,\cdots,\dperp\}$ are coordinate-axis indices.  The
first derivative obviously vanishes when evaluated at $\vec{r}_0$
because of a reflection symmetry of the lattice, showing that the net
force on the antibrane at this point is zero.

It is perhaps more remarkable that the second derivative also vanishes
at $\vec{r}_0$. The vanishing of the leading contribution may be
verified by explicit summation of the $2^\dperp$ nearest
neighbors. More generally, the vanishing of the entire sum may be seen
as follows. Consider the two cases, $a = b$ and $a \ne b$:
\begin{enumerate}
\item When $a=b$, given any lattice point $\vec{r}_1$, we can generate
other lattice sites equidistant from $\vec{r}_0$ (if necessary, we
choose a coordinate system such that $\vec{r}_0$ is the origin, with
coordinate axes aligned along those of the lattice) by cyclically
permuting the coordinates of $\vec{r}_1-\vec{r}_0$.  We can generate
at most $\dperp-1$ new points in this way.  If all these $\dperp$
points (including $\vec{r}_1$) are distinct, then
$\sum_{i=1}^\dperp(r_0-r_i)_a^2=|{\vec r}_0-{\vec r}_1|^2$ for all
$a$, where the summation is over these lattice sites.  If instead only
$\ell<\dperp$ points are distinct (so that they transform under a
${\bf Z}_{\ell}$ subgroup of the ${\bf Z}_{\dperp}$ group of cyclic
permutations of the coordinates), then Lagrange's factor theorem for
the dimension of a subgroup implies that $\dperp\sum_{i=1}^\ell
(r_0-r_i)_a^2/ |{\vec r}_0-{\vec r}_1|^2=\ell$ for all $a$, which
cancels the contribution of the $\delta_{ab}$ term in Eq.~\pref{VDD}.
\item When $a\neq b$, the second derivative vanishes due to a
reflection symmetry along any plane where, say, $(r_0-r_i)_a$ is a
constant.
\end{enumerate}

\EPSFIGURE[r]{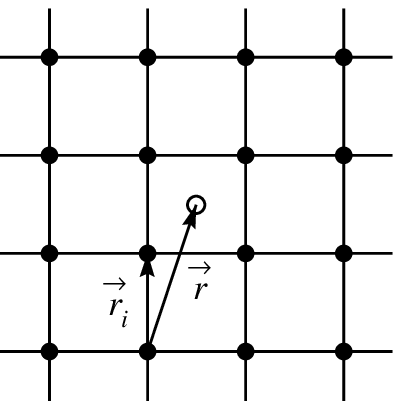,width=6cm}
{The covering space of a square torus. Solid circles represent
images of the $Dp$ brane, and the open circle represents the
$\overline{Dp}$ brane, which is near the center of a hyper-cubic cell
during the slow-roll inflationary phase. The images of the antibrane
are omitted because they do not contribute to the potential of the
antibrane since the force between an antibrane and another antibrane
is zero.\label{figure1}}

We therefore conclude the second derivative of the potential is zero
along any direction at this special point of the hyper-cubic cell.
Of course, this conclusion depends upon our choice of the square lattice.

Consider now antibrane motion when the antibrane is near the center of
a hyper-cubic cell. In this case we may expand the potential in terms
of power series of this small displacement $\vec z$ from
the center. From a simple symmetry argument, one can easily see that
all the odd powers in $\vec z$ vanish.  Furthermore, we have shown
above that the quadratic terms in $\vec z$ also vanish, so that the
leading contribution to the potential is the quartic term in $\vec z$.

We can therefore model the relative motion of the branes in a
quartic potential
\begin{equation}
V(z)\,=\,A-{1\over4}\,C\,z^4\,,
\end{equation}
where $A$ is defined as before and
\begin{equation}
C=\gamma\,\ms^{-(2+d)}\,e^{2\phi}\,T^2_p\,V_\parallel\,
r_\perp^{-(2+\dperp)}
\end{equation}
with $\gamma$ a constant of order ${\cal
O}(1)$.  It is straightforward to derive for this potential the slow
roll parameter $\eta$ and the density perturbation $\delta_H$ 
\eq 
\eta \approx -3\,\gamma \left({z\over\rp}\right)^2 \,, \qquad\qquad
\delta_H \approx {2\over5\,\pi}\,\sqrt{\,\gamma\over3}\, {N^{3/2}
\over \mpl\,\rp} \,, 
\eeq 
where again we have used the standard slow-roll equations. We see now that
slow-roll is guaranteed for sufficiently small $z$.

The value of the antibrane displacement from the center of the
hyper-cubic cell at the time when the COBE perturbation leaves the
horizon during the inflation is
\begin{equation}
z_*\approx{\rp\over\sqrt{\,2\,\gamma\,N}}\,.
\end{equation}
From this the spectral index and its derivative in this model can be
easily determined
\begin{equation}
n-1 \approx -\,{3\over N}\,,\qquad\qquad
{\rmd n\over\rmd\ln k} \approx -\,{3\over N^2}\,.
\end{equation}
One can also compute similar quantities for primordial tensor
perturbations.

Using $N \approx 60$ we find $z_*\approx 0.1\,\rp$ and the value of
the displacement at the end of slow roll $z_{\rm end}\approx 0.5\,\rp$
(from $|\eta|\approx 1$). Both of these are fairly natural numbers.
Comparing to the COBE normalization~\cite{COBE}, $\delta_H\approx
1.9\times 10^{-5}$, we obtain the size of the compactified transverse
dimension
\begin{equation}
\mpl\,\rp\approx 1.3\times 10^6 \qquad {\rm or}\qquad
r_\perp^{-1}\approx 10^{12}~{\rm GeV}\,.
\end{equation}
The deviation of the spectral index $n$ from the scale-invariant form
is at a few percent level, and the spectrum is slightly red.  $\rmd
n/\rmd\ln k$ is at the order of ${\cal O}(10^{-3})$.  These features
of the predicted CMB spectrum are quite common for inflationary
models, and are consistent with current data~\cite{cmbexp}. They will
be tested more precisely in future satellite experiments such as MAP
and Planck.

\section{Collisions, Reheating and Cascade}

In the final stages of brane-antibrane annihilation, many more degrees
of freedom are relevant than simply the inter-brane separation.  This
is a good thing, since these provide many channels for reheating which
are not constrained by the flatness of the potential responsible for
the initial slow roll of the inflationary phase. In this general sense
brane collisions are an extreme example of hybrid
inflation~\cite{hybridinflation}.

The connection to hybrid inflation can be made much more explicit
within string theory. The amplitude for brane-antibrane annihilation
develops a negative mode when the distance between a brane and
antibrane shrinks to $M_s^{-1}$.  This signifies the opening up of new
channels for annihilation. In particular, it signals the appearance
of tachyonic open strings connecting the branes to the antibranes
\cite{Tachyon, Banks}. The mass of the
tachyon is a function of the inter-brane distance and is given
by~\cite{Polchinski}\
$M_T^2 = y^2/(2\pi\alpha')^2 - 2/\alpha'$. Here, $y$ is the inter-brane
separation and $\alpha' \sim \ms^{-2}$.  Clearly this mass is
tachyonic for small $y$, changing sign when $\ms\,y \gsim 1$.

Although the complete form of the potential as a function of both $y$
and $T$ is not known, the limiting forms for $T=0$ and $y=0$ are. For
large $y$, the large positive $M_T^2$ localizes $T$ at zero, leaving
only the relatively flat function of $y$ on which the analysis of the
previous sections was based. More recent work has also given the
potential for $y=0$, in which case the two-derivative truncation of
the exact tachyon action (in the dimensionally-reduced form) is
conjectured to be~\cite{tseytlin}
\eq
\label{Taction}
S_T = -\ms^2\,\mpl^2 
\int\rmd^4x\,e^{-|T|^2}\,\left[\,1 + \kappa_1 
\left( {2 + \kappa_2\,|T|^2 \over \ms^2} \right) |\partial T|^2\,
\right]\,.
\eeq
Here $\kappa_1$ and $\kappa_2$ are calculable, dimensionless 
${\cal O}(1)$ numbers, whose values need not concern us further.
Only one feature of Eq.~\pref{Taction} is important
for our present purposes: the qualitative shape of the potential well,
which is a Mexican hat potential once expressed in terms of the canonically
normalized scalar field.

Thus, the tachyon potential behaves as follows: while the branes are
still far apart, the potential is parabolic and the tachyon field is
localized at the bottom of the potential, $T=0$.  Once the distance
shrinks to less than $M_s^{-1}$, the potential develops a double well
shape and the tachyon starts to roll downwards until it reaches one
of the degenerate minima of the potential.  If the winding number of
the tachyon is zero, then the brane and antibrane will annihilate.
According to a famous conjecture of Sen~\cite{sen}, the height of the
tachyon potential is equal to the brane tension and 
the end product of brane-antibrane annihilation is the closed string
vacuum. On the other hand, if the
tachyon has non-trivial winding, then stable lower dimensional
branes will be formed after it condenses.

Many types of lower dimensional branes may be
formed~\cite{lowerbranes}. For example, on
the worldvolume of the $Dp$--$\overline {Dp}$ pair there are kink-like
solutions to the tachyon field equations that correspond to unstable
non-BPS $D(p-1)$ branes. These $D$-branes in turn contain a real
tachyon field which induces them to decay into stable BPS $D (p-2)$
branes.  Equivalently, these $D(p-2)$ branes correspond to vortex
configurations of the complex tachyon that lives on the worldvolume
of the $Dp$--$\overline {Dp}$.  By
considering more than just a single brane-antibrane pair it is
actually possible to describe a system of $D(p-2)$ branes. Furthermore, by
looking at anti-kink solitons on the unstable
$D(p-1)$ branes one obtains $\overline{D(p-2)}$ branes. This then allows the
formation of $(p-2)$-dimensional brane-antibrane systems that can
lead, in turn, to $D (p-4)$ branes and so~on. 
Starting from a system of $D9$--$\overline{D9}$ branes we can thus generate all
possible $D$-brane configurations in Type I and Type IIB string theory.
A similar story holds for Type IIA, with the non-BPS $D9$ brane as the
starting point. This is the basis of the $K$-theory classification of
$D$-branes~\cite{k-theory}.

The tachyon potential above may have important applications for
cosmology.  We have checked that in its simplest setting it does not
produce slow-roll inflation by itself, as might be expected given
the absence of small parameters.  Its role begins once inflation
ends, when the inter-brane separation becomes of order $1/\ms$. Once
this point is reached, the tachyon mass changes sign and $T$ is free
to quickly roll away from zero.

In a field theory example~\cite{linde}, this roll along the tachyonic
direction has been shown to be very rapid, with the initial
potential energy being quickly converted into gradient 
energy for the tachyon field, and may also result in topologically
non-trivial configurations. We regard the quickness of this transfer of
energy into gradients of $T$ as a sign that a significant number of
lower-dimensional daughter branes are produced in each brane-antibrane
collision. 
If most of these daughter branes very quickly find nearby
antibranes to annihilate, liberating the gradient energy to heat, then
we are left with a post-collision world involving some remnant
lower-dimensional branes, and with much of the collision energy
converted into exciting both bulk modes and the internal degrees of
freedom on the daughter branes. A crude estimate of the reheat
temperature is given by equating $T_{\rm rh}^4$ with the initial
potential energy density, leading to
\eq
\label{Treh}
T_{\rm rh} \sim \left[ {2\alpha e^{\phi} \over (\ms\rp)^\dperp}
\right]^{1/4} \sqrt{\ms\,\mpl} = \left[ { 2\alpha e^{\phi} \over 
(\mpl\rp)^\dperp} \; \left({\mpl \over \ms}\right)^{\dperp-2}
\right]^{1/4} \mpl\,. 
\eeq
For instance if $d=6$ and $p=5$ and we take $\mpl\,\rp \sim 10^6$ (as
is suggested by the amplitude of primordial density fluctuations) and
$\alpha\,e^{\phi} \sim 10^{-2}$ and $\mpl/\ms \sim 10^3$ then we find
$T_{\rm rh} \sim 10^{13}$ GeV.

Although a careful analysis of this collision process is beyond the
scope of this paper, we remark that the picture has several attractive
features:
\begin{itemize}
\item
Because the collisions can have daughter branes, the reheating process
can be expected to efficiently excite the brane modes as well as those
of the bulk, as is required if most of the observable universe now
lives on one of these branes.
\item
Since the collision process produces daughter branes of lower
dimension, successive episodes of brane collisions will move the brane
population away from the higher-dimensional branes. Furthermore, since
lower dimensional branes have more difficulty finding one another than
do higher dimensional ones, this annihilation process will eventually
be self-limiting. The branes found in the late universe will have a
dimension which is small enough to preclude their having found
anti-partners to annihilate during the history of the universe.
\item
There is a simple estimate -- following an argument similar to that of 
Ref.~\cite{BV} -- of what dimension a brane must have in order to
plausibly fail to find partners with which to annihilate, and so to
dominantly survive into the present.  The basic
observation is that $p$-branes---whose world-sheets are
$(p+1)$-dimensional---generically have difficulty missing one another
if they are moving within a spacetime whose dimension is less than or
equal to $2(p+1)$. For 10-dimensional spacetime, passing $p$-branes
with $p\ge 4$ will generically collide, while those with $p\le 3$ will
generically miss.  We see then that if we start with a 10-dimensional
spacetime only branes of dimensions greater or equal than 3 have
finite probability of colliding with each other. If we start with
a system of $D9$--$\overline{D9}$ branes they will give rise to pairs
of $D7$--$\overline{D7}$ which can still annihilate each other to
produce pairs of $D5$--$\overline{D5}$ and then
$D3$--$\overline{D3}$. However, assuming the 3-branes can resolve the
compact dimensions,
these 3-branes will have negligible probability of finding
each other.  This implies that the cascade naturally stops there.
\item
Furthermore, as is illustrated in Fig. 2, brane collisions can 
plausibly act to align the surviving
branes parallel to the noncompact dimensions of space. This is because
collisions between branes wrapped around the compact dimensions and
branes parallel to the noncompact dimensions tend to leave daughter
branes which are both parallel to the noncompact dimensions and wrapped
around the compact ones. Furthermore, the residual wrapping about the 
extra dimensions can be removed through the collisions of branes with
opposite winding number, such as discussed in Ref.~\cite{BV}.
\item
Within the cascade picture there are
many branes colliding, but only some will cause inflation: those which
move slowly enough to give sufficient $e$-foldings. Once inflation
happens, the inflated area dominates the volume of the universe. 
For branes wrapped about a common direction there is an energetic cost 
to be not straight and parallel with one another, so given enough time
branes will tend to align and straighten.
Now one of the reasons it is hard to find real string potentials which 
inflate is that, in general, adjustments of brane energy (like
removal of wiggles, or alignment of branes, or their relative rotations)
take place on string timescales, making them too fast for inflation.
But since inflation preferentially happens for the slowest collisions,
it is plausible that these other processes will have had time to already 
have taken place, tending to make the branes align, be straight, etc. So 
we believe that the simplifying assumptions we make for our initial
conditions are plausible for those relatively rare collisions which
are of most interest for inflation. This plausibility
needs to be further investigated within our cascade scenario.
\end{itemize}

\EPSFIGURE[c]{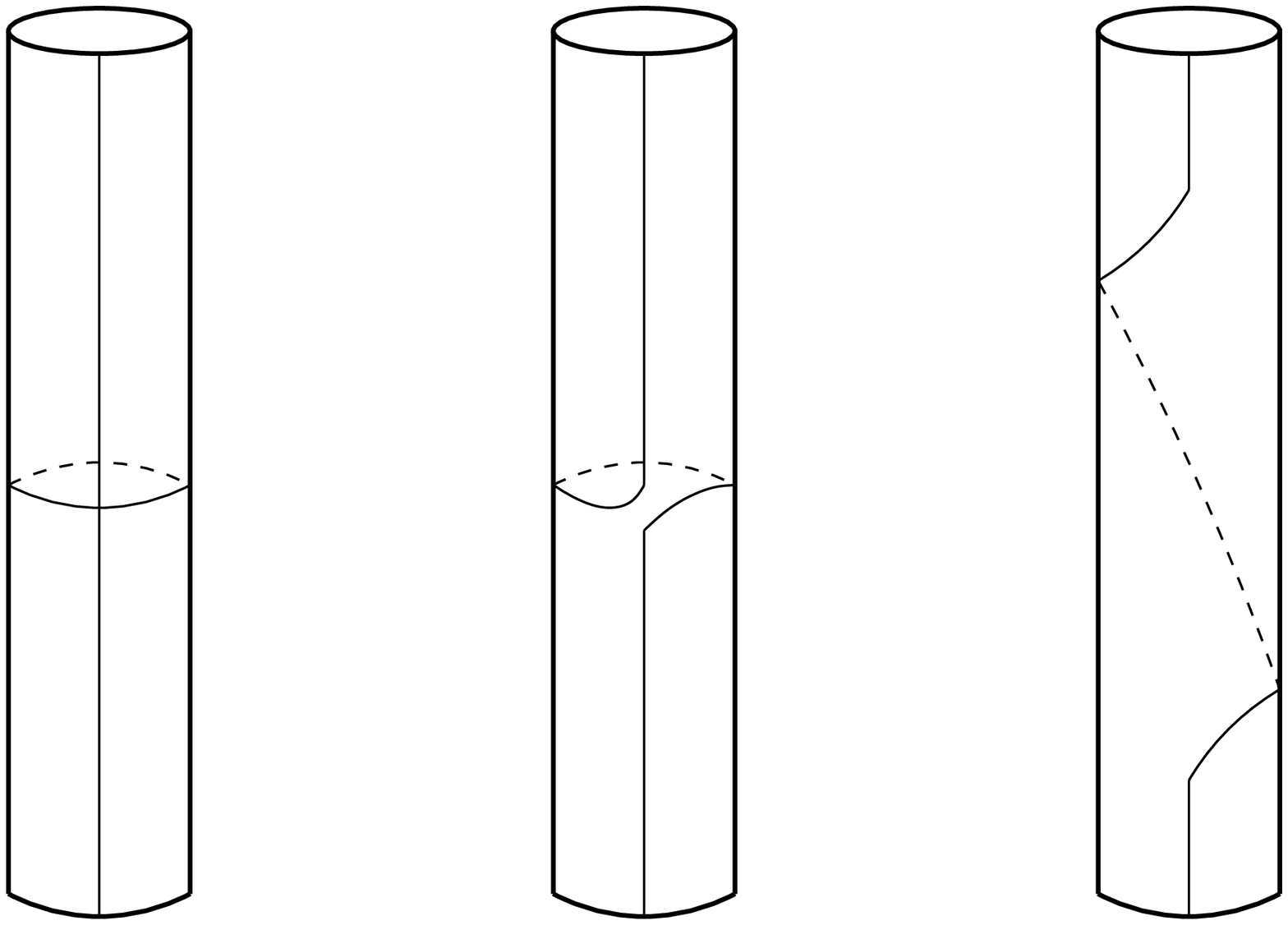, width=7cm}
{The figure on the left shows two branes along orthogonal
directions of a long cylinder. In the middle, the branes reconnect. The
final configuration, shown on the right, consists of a single brane
winding once around the compact direction.
\label{figure2}}   

It is quite intriguing that a $D3$ brane universe with branes
aligned with the large dimensions of space seems to be selected
in such a natural way. Notice that our arguments are not
identical to the one given in~\cite{BV}, but complements it in an
interesting way.  Their argument was based on the fact that winding
modes tend to compactify the space dimensions where they are winding
around and then prefer only three spatial dimensions to become large
(in the original counting, winding strings with $p=1$ cannot annihilate
in dimensions larger than $4=2(p+1)$), therefore splitting naturally
the ten dimensional spacetime into four large and six small
dimensions. Our argument would complement this by stating that in this
spacetime there will be stable $D3$ brane-worlds where we may
live.\footnote{More specifically, one can imagine the following
scenario: Branes of all types are initially there, and all dimensions
are compact but small. The large dimension branes annilate out 
instantly, leading a population of branes which dominantly involves
3-branes or lower. The winding of these various branes keeps any
dimensions from growing. Then the Brandenberger-Vafa mechanism kicks
in, making 4 dimensions large and 6 small, with no windings about
the large spatial directions. Energetics arguments can prevent 
these small dimensions from growing or collapsing. After this,
we have the particular collision which causes inflation of the
large 4 dimensions, as studied in the previous section.}
Notice that in some sense the assumptions of our argument are
less constrained than that in the discussion of~\cite{BV}. For their
argument to hold they have to assume all the spatial dimensions to be
toroidal, in our case the argument applies independent of the
properties of the full ten-dimensional spacetime.\footnote{Actually,
the $D3$ brane-world is preferred if we start with type IIB or type I
strings for which the starting point in the cascade are pairs of
$D9$--$\overline{D9}$ branes. In type IIA, the
starting point would be non-BPS $D9$ branes which then would give rise
to even dimensionality BPS $D$ branes and the cascade would stop with
4-branes.}

We see that a cosmology of cascading brane annihilations may explain our
finding ourselves on a 3-brane parallel to the large dimensions,
by relating it to the prediction that spacetime is ultimately
10- or 11-dimen\-sional \footnote{On the remaining 3-brane we may have 
remnant cosmic strings and monopoles which could overclose the Universe.
In order to properly pose and solve this question a more detailed
knowledge of the final state distribution of branes after annihilation
is required, including a good understanding of what the branching fraction
for the production of daughter branes (as opposed to no branes)
given a brane-antibrane collision. Also an understanding of how these
daughters are distributed given the geometry of the initial brane/antibrane
collision. These considerations  are beyond
the scope of our paper (see the second article of reference \cite{BV}\
for a related discussion). }.  Although the arguments we propose along
these lines are preliminary and qualitative, we believe they have
sufficient promise to deserve more careful further
investigation.

Let us finish by recapitulating the results of this article.
In the first part of our paper we asked the question: Is the
brane/antibrane potential leading naturally to inflation? Our answer
was no. We then asked
the question: Is inflation possible at all in this system? and the answer
to this is yes, but only in special configurations when the potential
happens to be flat enough. This is the first time that inflation is achieved 
from a computable potential in string theory. Furthermore the, also computable,
stringy tachyon potential provides the natural way to exit inflation 
in an elegant  realization of hybrid inflation in terms of 
 string theoretical fields. The fact that at the minimum of the tachyon
potential supersymmetry is recovered guarantees the vanishing of the
 vacuum energy at that point without any fine-tunning.

Our proposal of the cascade mechanism, obtained from the  
tachyon potential, then  addresses
the issue of how to accomplish inflation naturally if it is so hard to
achieve. If only two branes exist, it would be fine-tuned to expect
them to collide in the way which inflates. But in the cascade picture
the universe starts in a generic state in which many branes are
present and colliding, and so the inflationary collisions are rare, but
happen. Furthermore we can trace back the origin of the branes gas to 
overlapping, static, 9-brane/anti 9-brane systems as providing the initial
 conditions for the multi-brane Universe. We believe our scenario has
many interesting features deserving further investigation.

\acknowledgments{
We would like to thank A. Aguirre, V. Barger, K. Benakli, S. Gratton,
C. Grojean, T. Han,
J. Khoury, E. Kiritsis, M. Klein, J. March-Russell, D. Mateos, G. Ross,
A. Tseytlin, N. Turok, E. Zavala and B. Zwiebach for useful
conversations and C. R. Burgess and E. Quevedo for their assistance. 
M.M. thanks A. Davis for various discussions. The
work of CPB was supported in part by N.S.E.R.C. (Canada),
F.C.A.R. (Qu\'ebec) and the Ambrose Monnell Foundation.  DN was supported in
part by DOE grant DE-FG02-90ER40542.  MM was partially supported by
the Isaac Newton Institute and FQ by PPARC. GR was supported in
part by NSF grant NSF-PHY-0070928 as well as a Helen and Martin
Chooljian fellowship and RJZ by NSF grant NSF-PHY-0070928. 
}

\vskip 0.2in 
{\noindent\it Note added}: 
The results of this article
were presented at the conference ``From the Planck Scale to the Weak
Scale'' Les-Londes-de Maures, France. We learned there of related work
by G. Dvali, S. Solganik and Q. Shafi~\cite{DvaliShafi} that was also
presented in the conference and overlaps with the content of Sections
2.1--2.4, however, we disagree with their conclusion (see 
Footnote \ref{footnote}). 
We thank them for bringing this work to our attention.


\end{document}